\documentclass[aps,prl,twocolumn,superscriptaddress]{revtex4-2}
\usepackage{graphicx}
\usepackage{amsfonts}
\usepackage{amsmath}
\usepackage{amssymb}
\usepackage{bm}
\usepackage{hyperref}
\usepackage{slashed}
\usepackage{color}
\usepackage{ulem}

\begin{document}
\title{Spin-charge Kondo effect for a quantum dot with side coupled Majorana zero mode}
\author{Haojie Shen}
\affiliation{School of Physics and Astronomy, Shanghai Jiao Tong University, Shanghai 200240, China}

\author{Wei Su}
\email{suwei@sicnu.edu.cn}
\affiliation{College of Physics and Electronic Engineering, Center for Computational Sciences, Sichuan Normal University, Chengdu 610068, China}
\affiliation{Beijing Computational Science Research Center, Beijing 100084, China}

\author{Mengnan Chen}
\affiliation{School of Science, Hangzhou Dianzi University, Hangzhou 310027, China}

\author{Xiaoqun Wang}
\email{xiaoqunwang@sjtu.edu.cn}
\affiliation{School of Physics, Zhejiang University, Hangzhou 310027, China}

\date{\today}

\begin{abstract}
We investigate a minimal system consisting of a quantum dot coupled to a Majorana zero mode and a normal lead. We identify the underlying screening process as a novel spin-charge Kondo effect, where the low-energy spin and charge degrees of freedom of the Majorana zero mode-quantum dot subsystem are fully screened by those in the normal lead, resulting in the formation of a spin-charge singlet. An effective low-energy model is derived, with charge fluctuations appropriately accounted for. This spin-charge Kondo effect is found to be consistent with the spin-dependent Andreev/normal boundary conditions induced by the Majorana zero mode. We demonstrate that the anomalous substructure in the spectrum and thermodynamic properties is closely tied to the proportion of the charge component in the screening cloud. The spin-charge screening cloud exhibits scaling behavior analogous to that of traditional Kondo systems, though the sub-leading even-odd effect is subtly modified by the boundary conditions. These findings enhance our understanding of Kondo physics and resolve key debates on quantum dot nanostructures with Majorana zero modes.
\end{abstract}

\maketitle

\textcolor{blue}{Introduction}--
The Kondo effect, which arises from the interaction between a localized spin and a conduction electron sea, serves as a cornerstone for understanding many-body phenomena in strongly correlated electron systems\cite{Hewson1993}. Although spin-flip scattering plays a crucial role in high-energy regimes, below the Kondo temperature $T_K$, the low-energy physics can be understood as a delocalized spin singlet state that dynamically forms between the localized spin and the conduction electron spins in the bath. The Kondo screening cloud, which extends to a mesoscopic length scale, has only been observed in nanostructures in recent years\cite{VBorzenets2020}. After decades of study, Kondo-like systems involving composite impurities or unconventional baths remain an area of significant theoretical and experimental interest. Notable examples include the charge Kondo effect \cite{Taraphder1991,JarilloHerrero2005}, topological Kondo effect \cite{Cooper2012,Li2023,Li2023b}, Yu-Shiba-Rusinov states in superconductors\cite{Yu1965,Farinacci2020,Chiu2022}, mesoscopic nanostructures on surfaces \cite{Teri2000,Klochan2011}, and the emergent Kondo effect in spin liquids \cite{Wang2021}.

\begin{figure}
    \centering
    \includegraphics[width=0.95\linewidth]{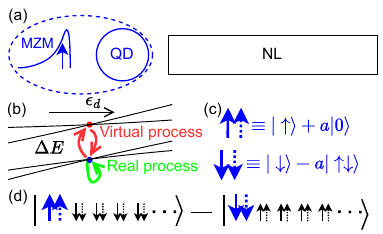}
\caption{(a)(b)(c) and (d) the schematic plot of the MZM-QD-NL system, the spectrum of the MZM-QD subsystem, low energy $SU_L(2)$ doublet for the MZM-QD subsystem, and the SC singlet screening cloud, respectively. The blue(black) double arrows denote SC doublet in the MZM-QD subsystem (NL).}
    \label{fig1}
\end{figure}

In the past decade, Majorana zero modes (MZMs) in topological superconductors (TSCs) have been at the forefront of research, owing to their potential applications in fault-tolerant topological quantum computing\cite{Read2000,Kitaev2001,Nayak2008,Fu2008}. When detecting or manipulating in tunneling transport setups\cite{Mourik2012,Das2012,Leijnse2011,Leijnse2011b}, the MZM-quantum dot(QD)-normal lead(NL)system(see Fig.\ref{fig1}(a)) is naturally formed. The MZM-QD-NL system were investigated extensively through perturbation theory, renormalization group and NRG calculations\cite{Lee2013,ChengMeng2014,Silva2020,Vernek2014,Cifuentes2019,Ramos2020,Piotr2022}. A new $A\otimes N$ fixed point was identified \cite{ChengMeng2014} and novel phenomenons were predicted. However, the fundamental question of whether MZM physics and Kondo physics are competitive, cooperative, or coexist is still under debate\cite{Lee2013,ChengMeng2014,Silva2020}. Unlike normal composite impurities, the MZM-QD coupling introduces a spin-polarized pairing channel to the QD, in addition to the hopping channel. One can thus expect that both spin and charge fluctuations are crucial in determining the low-energy physics. Furthermore, the half-Fermi nature of the MZM significantly alters the zero-temperature thermodynamic and spectral behavior. Due to the involvement of multiple mechanisms, it is difficult to obtain a complete understanding by focusing on only a subset of the observables.

In this letter, we treat the MZM-QD subsystem as the appropriate atomic limit, with the MZM-QD coupling being considered on equal footing with the Hubbard  $U$ at the QD. We find that it is the spin-charge (SC) degree of freedom that couples to those in the NL in the low-energy effective model. Multiple thermodynamic properties and the microscopic screening cloud were calculated using the numerical renormalization group (NRG) \cite{Wilson1975,Bulla2008,Weichselbaum2012} or density matrix renormalization group (DMRG) \cite{White1992,Schollwoeck2011} with  $SU(2)$  symmetry. We identify the underlying screening process as a novel SC one, where the low-energy SC degree of freedom in the MZM-QD subsystem is screened by those in the bath, as schematically shown in Fig.\ref{fig1}. The current SC Kondo effect differs from the traditional spin or charge Kondo effects in many aspects. In SC Kondo effect, charge fluctuations are inevitably present at low energy due to the MZM-QD coupling, irrespective of the value or sign of $U$. Furthermore, the low-energy characteristics induced by the half-Fermi nature of the MZM are closely related to the proportion of the charge component in the screening cloud. We also show that the SC screening cloud collapses onto a universal curve after subtracting the characteristic length. Within this framework, the previous interpretations of competition, coexistence, and cooperation are consistent, as they highlight different aspects of the phenomenon. The current SC Kondo effects are uniquely induced by the specific form of the MZM-QD coupling and are crucial for connecting theoretical predictions with experimental measurements.

\textcolor{blue}{\textit{Model and Atomic limit}--}The MZM-QD-NL system can be described  by the following Hamiltonian,
\begin{equation}
    H = H_{\mathrm{MQ}} + H_{\mathrm{NL}} + H_\textrm{hyb},
\end{equation}
Here $H_{\mathrm{NL}} = \sum_{\sigma \mathbf{k}}\epsilon_\mathbf{k} c^\dagger_{\mathbf{k}\sigma} c_{\mathbf{k}\sigma}$ is the Hamiltonian of the NL and $H_\textrm{hyb}= \sum_{\sigma\mathbf{k}} V_\mathbf{k} c^\dagger_{\mathbf{k}\sigma} d_{\sigma} + h.c.$ describes the QD-NL hybridization. In the following, we assume $V_\mathbf{k} = V$ as a constant. We consider an Anderson type QD coupled to one MZM with the Hamiltonian\cite{ChengMeng2014}
\begin{equation}
    \hat{H}_{\text {MQ }}=\sum_{\sigma} \varepsilon_{d} d_{\sigma}^{\dagger} d_{\sigma}+U n_{\uparrow} n_{\downarrow} +\lambda \gamma\left(d_{\uparrow }-d_{\uparrow}^{\dagger}\right)\label{model1},
\end{equation}
where $\gamma \equiv (f + f^\dagger)$ is the MZM operator at the topological defect of the TSC while $\lambda$ measures the MZM-QD coupling strength. Here, we assume that the pairing gap in the TSC is infinitely large relative to the energy scale of interest, and that the MZMs are well seperated such that the coupling between them are negligible.

The MZM-QD coupling generally breaks both the total spin rotation $SU_S(2)$ symmmetry and total charge rotation $SU_C(2)$ symmetry. At the symmetric point $U=-2\epsilon_d$, however, the total $SU_L(2)$ symmetry generated by $\mathbf{L} \equiv \sum_i \mathbf{l}_i$ remains conserved. Here $ \mathbf{l}_i\equiv  \mathbf{s}_i +  \bm{\eta}_i$, with $\mathbf{s}_i$ and $\bm{\eta}_i$ generator of the spin and charge rotation symmetry at site $i$, respectively[cite supplemental]. The total $SU_L(2)$ symmetry relies on the particle-hole symmetry in the NL. This requirement is generally satisfied, as only states near the Fermi surface are relevant in impurity physics. In the following, we will focus on systems with total $SU_L(2)$ symmetry for the sake of simplicity. However, slight deviation from this symmetry will not significantly alters the underlying physics. In addition, the total parity $P$ for the number of fermions is conserved, and there always exists a degeneracy between all states with $P=0$ and $P=1$, reflecting the topological nature of the MZM.

The energy spectrum of the MZM-QD subsystem already provides significant insights. When varing $\epsilon_d$ at the QD with $U$ and $\lambda$ fixed, an energy gap $\Delta E$ well separates the low energy states (denoted by $|l\rangle$s) from the high energy ones (denoted by $|h\rangle$s), as shown schematically in Fig.\ref{fig1}(b). Exactly at the symmetric point $U=-2\epsilon_d$, the four states in $|l\rangle$s ($|h\rangle$s) can be grouped according to their parity, i.e., one group with $P=0$ and the other $P=1$. Two states within each group then form a $SU_L(2)$ doublet with $l^z = \pm 1/2$. When $\lambda=0$, the $SU_L(2)$ doublet reduces to the singly occupied $SU_S(2)$ one as in traditional spin Kondo systems (up to a topological degeneracy due to the uncoupled MZM). As $\lambda$ grows, the contribution of empty $|0\rangle$ and the doublely occupied configuration $|\uparrow\downarrow\rangle$ increase correspondingly. When coupling to the the NL, an $SU_L(2)$ singlet is formed between the local states and states in the NL as scheched in Fig.\ref{fig1}(d). The SC degrees of freedom at the MZM-QD subsystem will be fully screened by the SC cloud in the NL.

\begin{figure}
    \centering
    \includegraphics[width=1.0
    \linewidth]{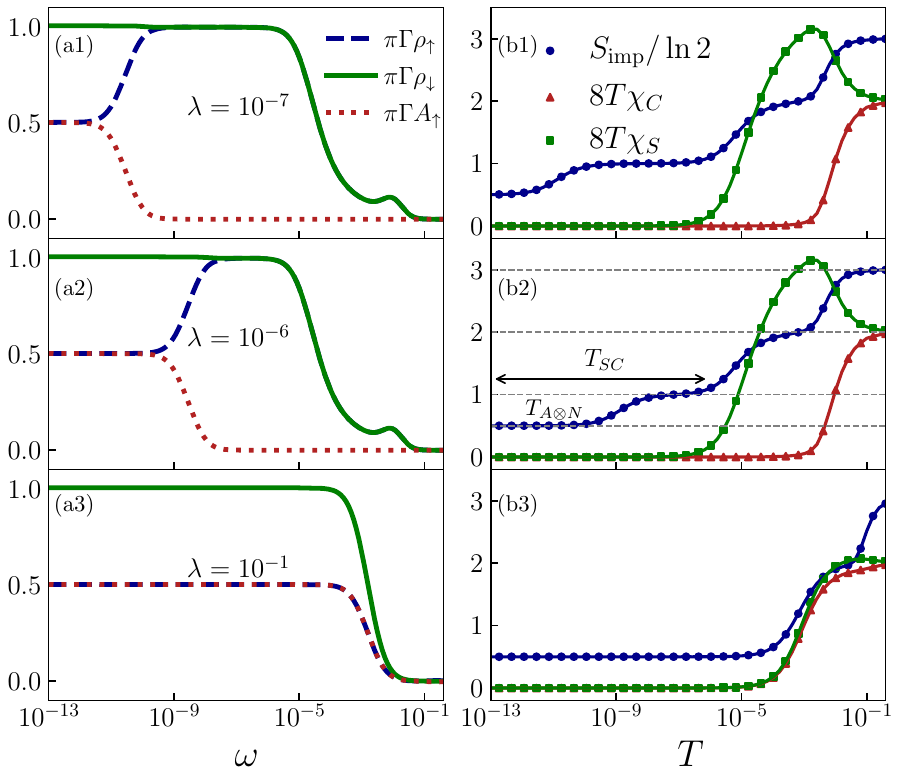}
\caption{Left panel: LDOS for $\rho_\uparrow$, $A_\uparrow$ and $\rho_\downarrow$  of various $\lambda$s calculated with NRG. Right panel: The corresponding MZM-QD entropy $S_\textrm{imp}$, charge susceptibility $\chi_C(T)$ and spin susceptibility $\chi_S(T)$. Other parameters are $\Gamma=0.0016$, $\Lambda=2.5$ and $U=-2\epsilon_{d}=0.02$.
    }
    \label{fig2}
\end{figure}

\textcolor{blue}{\textit{Spectral functions and thermal properties}--}
The spectral function is defined as the imaginary part of the retarted Green's function $\langle\langle \hat{A}| \hat{B} \rangle\rangle_\omega$, which can be accurately calculated through NRG within the framework of FDM\cite{Anders2005,Weichselbaum2012}. By setting $\hat{A}$($\hat{B}$) to $d_\sigma$ ($d^\dagger_\sigma$) we get the normal density of states(DOS) $\rho_{\sigma}(\omega)$ at the QD. When the MZM is decoupled from the QD ($\lambda=0$), the DOS exhibits the well-known Kondo resonance peak of width $T_K$ centered at $\omega \sim 0$. The MZM-QD coupling gives rise to an anomalous DOS component $A_{\uparrow}(\omega)$ with both $\hat{A}$ and $\hat{B}$ being $d_\uparrow$. Fig.\ref{fig2} (a1-a3) show the spin-resolved (anomalous) DOS at the QD for various $\lambda$s. A tiny $\lambda$ modifies $\rho_\uparrow$ by inducing a dip of depth $0.5$ near $\omega\sim 0$ through anti-Fano resonance\cite{Lee2013}, leaving $\rho_\downarrow$ unaffected[see Fig.\ref{fig2} (a1-a2)]. Correspondingly, $A_\uparrow(\omega)$ has a resonance peak of height $0.5$, whose width $T_{A\otimes N}$ is the same as that of the dip. Actually, one can prove that $\rho_\downarrow=\rho_\uparrow + A_\uparrow$ because of the $SU(2)_L$ symmetry. The width of the dip in $\rho_\uparrow$ (and $T_{A\otimes N}$) grows progressively when increasing $\lambda$ until $T_{A\otimes N}\sim T_K$, after which the dip spreads throughout the resonance peak in $\rho_\uparrow$ and the width of $\rho_\downarrow$ start to grow. Finally, for a sufficiently large $\lambda$, one has $\rho_\uparrow(\omega) = A_\uparrow(\omega) = \frac{1}{2} \rho_\downarrow(\omega)$. As will be discussed bellow, the width of the resonance peak in $\rho_\downarrow$ (defined with $T_{SC}$) can be regarded as a characteristic scale of the SC screening process. One has $\lim_{\lambda\to 0} T_{SC} = T_K$, i.e., as the MZM-QD coupling goes to zero, the SC screening process returns back to the traditional spin one.

The substructures in spectral function have clear correspondence in the MZM-QD entropy $S_{\textrm{imp}}$, which is defined as the entropy difference between the total system and the NL subsystem. Four plateaus in Fig.\ref{fig2}(b1-b3) can be identified as four fixed points. The three rightmost(high temperature) plateaus with $S_{\textrm{imp}} = 3\ln 2$, $2\ln2$ and $\ln 2$ correspond to the free, local moment, and strong coupling fixed points, respectively. A non-zero $\lambda$, however, modifies the local moment and the strong coupling fixed points to those in terms of the SC degree of freedom. A noteworthy result is that the SC strong coupling fixed point is unstable due to the MZM-QD coupling. As temperature decays to zero, the system finally flows to the $A\otimes N$ fixed point \cite{ChengMeng2014} with $S_{\textrm{imp}} = \ln 2/2$, which is characterized by a Andreev(normal) boundary condition for the spin-$\uparrow$($\downarrow$) electrons. For a large $\lambda$, however, the strong coupling fixed point is no longer accessible, and the system flows directly from the local moment fixed point to the $A\otimes N$ one as shown in Fig.\ref{fig2}(b3).

One of the main results of this work is that the SC degrees of freedom is fully screened in both the SC strong coupling and the $A\otimes N$ fixed points, and there is a adiabatic crossover from the spin Kondo effect to the SC one as $\lambda$ growing. By adding perturbation $h l^z_{\textrm{QD}}$ to the QD, the SC susceptibility can be obtained through $\chi_{SC}(T)=\chi_S(T)+\chi_C(T)$. Here $\chi_S(T)\equiv \partial\langle s^{z}_{\mathrm{QD}}\rangle/\partial h$ and $\chi_C(T)\equiv \partial\langle \eta^{z}_{\mathrm{QD}}\rangle/\partial h$ as $h\to 0$ are susceptibilities for the spin and charge, seperatively. For a tiny $\lambda$ with $T_{A\otimes N}\ll T_K$, $\chi_C(T)$ ($\chi_S(T)$) tends to zero when the system following to the local moment (strong coupling) fixed point as shown Fig.\ref{fig2}(b1-b2), resembles the behaves in spin Kondo effect. For large enough $\lambda$ with $T_{A\otimes N} \gtrsim T_K$, however, both $\chi_C(T)$ and $\chi_S(T)$ remain significant in the local moment region as shown in Fig.\ref{fig2}(b3), untill the $A\otimes N$ fixed point is reached.

\begin{figure}
    \centering
    \includegraphics[width=1.0\linewidth]{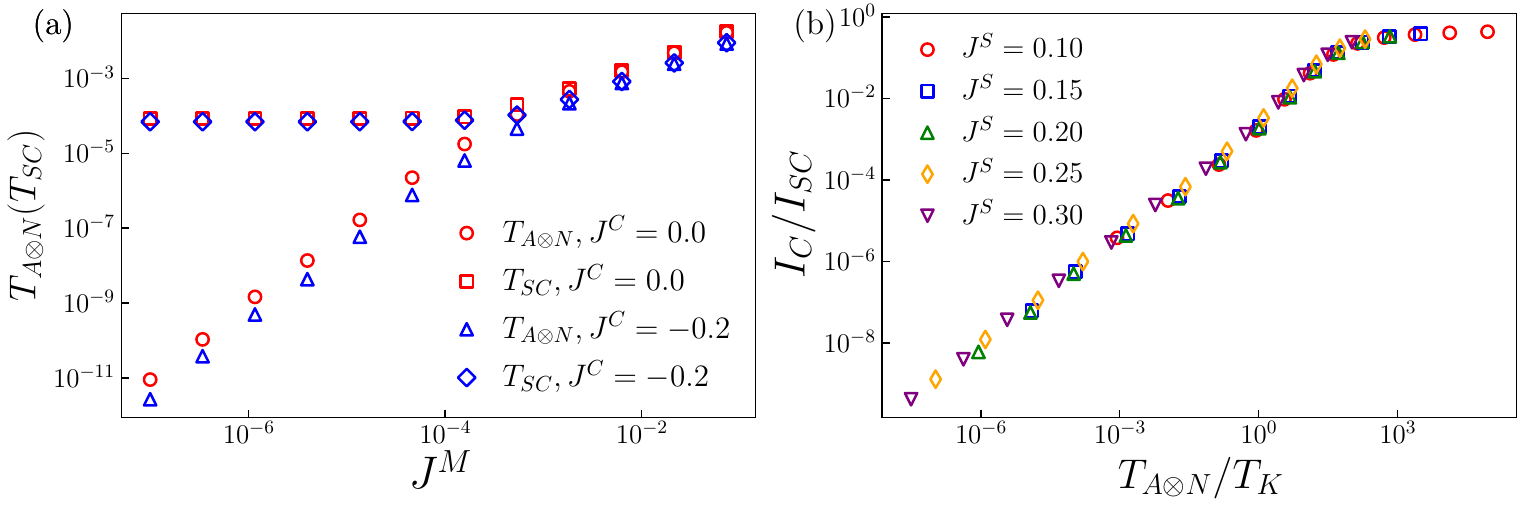}
\caption{(a) Characteristic temperature $T_{A \otimes N}(T_{SC})$ for the $A\otimes N$(SC strong coupling) fixed point as function of $J^M$ by fixing $J^{S} = 0.25$. (b) Contribution ratio $I_C/I_{SC}$ as a function of $T_{A \otimes N}/T_{K}$ for various $J^S$ and $J^C = 0$.}
    \label{fig3}
\end{figure}

\textcolor{blue}{\textit{Low energy effective model}--}
The low-energy effective model is obtained by performing the Schrieffer-Wolff transformation, i.e., by projecting out the high-energy states in the same sense of the spin Kondo effect. However, by taking the MZM-QD subsystem as the atomic limit, the low-energy states are $SU_L(2)$ doublets, comprising both spin and charge degrees of freedom. The projection is now performed by evaluating the inner products between wavefunctions of the MZM-QD subsystem[cite supplements]. The effective Hamiltonian is given by
\begin{equation}
H^J_{\textrm{eff}}=J^M(\mathrm{i}\gamma\mathbf{l}_{\textrm{QD}}/\theta)\cdot\bm{\xi}_{0}+J^S\mathbf{l}_{\textrm{QD}}\cdot\mathbf{s}_{0}+J^{C}\mathbf{l}_{\textrm{QD}}\cdot\bm{\eta}_{0}.
    \label{eq4}
\end{equation}
Here $\theta = -\epsilon_d / \Delta_E$ with $\Delta_E = \sqrt{\epsilon_{d}^{2}+4\lambda^{2}}$, and the subscript $0$ denotes the origin in the bath that couples to the QD. The coupling constants are given by $J^M = 2\lambda V/\Delta_E$,  $J^S = 2 V^2\epsilon_d(\epsilon_d-\Delta_E)/\Delta_E^3$, and $J^C = 2 V^2\epsilon_d(\epsilon_d+\Delta_E)/\Delta_E^3$, respectively. 

The $J^M$ term arises from $H_{ll} = P_l H P_l$ ($P_l \equiv |l\rangle \langle l|$), with the operator product given by $\mathrm{i}\gamma \mathbf{l}_{\textrm{QD}} \cdot \bm{\xi}_0 \equiv \gamma l^{z}_{\textrm{QD}} (c_{0\uparrow} - c_{0\uparrow}^{\dagger}) + \gamma l^{+}_{\textrm{QD}} (-c_{0\downarrow}^{\dagger}) + \gamma l^{-}_{\textrm{QD}} c_{0\downarrow}$. Here, the charge fluctuations transition states between the $P = 0$ and $P = 1$ groups and flip $l^z_{\textrm{QD}}$ within each low-energy doublet, as illustrated by the ``real process'' in Fig.\ref{fig1}(b). On the other hand, both the $J^S$ and the $J^C$ terms result from the ``virtual process'' (see Fig.\ref{fig1}(b)) mediated by the high-energy states. The former term exchanges the $SU_S(2)$ spin between the MZM-QD subsystem and the NL, analogous to the spin Kondo effect. The $J^C$ term couples the $SU_C(2)$ charge degree of freedom in the MZM-QD subsystem to those in the NL, which is also a consequence of the MZM-QD coupling as implicitly included in $\mathbf{l}_{\textrm{QD}}$. In the limit of $U\to \infty$, the charge fluctuation is freezed at the QD such that $\mathbf{l}_{\textrm{QD}}\to \mathbf{s}_{\textrm{QD}}$, $(\gamma\mathbf{l}_{\textrm{QD}}/\theta)\to \gamma\mathbf{s}_{\textrm{QD}}$, and the $J^C$ term is ommitable. This effective low-energy model reduces to that in Ref.\cite{ChengMeng2014}. The current effective model, however, is applicable for a much wilder paramenter region by giving that $V\ll \Delta_E$. It is valid even with $\lambda\gg U$ when $\lambda$ is large enough, i.e., far away from the Kondo limit. Furthermore, it is the SC degree of freedom at the MZM-QD subsystem that couples to those in the NL.

With the NL modeled as a Wilson chain of length $N$, the characteristic temperature $T_{A\otimes N}$ ($T_{SC}$) for the $A\otimes N$ (SC strong coupling) fixed point can be extracted from $S_{\textrm{imp}}$ as it reaches $0.99 \ln 2$ ($1.01 \ln 2$), as shown in the middle subplot of Fig.\ref{fig2}(b). In the limit of $J^M = 0$, $T_{SC}$ reduces to the spin Kondo temperature $T_K$. In Fig. \ref{fig3}(a), we show $T_{A\otimes N}$ and $T_{SC}$ as functions of $J^M$ with $J^S = 0.25$. For sufficiently small $J^M$, $T_{A\otimes N} \ll T_{SC}$ and $T_{SC} \sim T_K$ remains unchanged, meaning the spin degrees of freedom dominate the screening process. However, for large enough $J^M$, $T_{A\otimes N}$ converges with $T_{SC}$, and $T_{SC}$ becomes the only characteristic temperature, which increases linearly as $J^M$ increases. A moderate $J^Q$ term only slightly modifies the results and will therefore be omitted in the following discussion.

The screening cloud can be extracted from the correlation function $C(x) \equiv \langle \mathbf{l}_\textrm{QD} \cdot \mathbf{l}_x \rangle$, where $x$ is a position in the NL. By integrating $C(x)$ over all sites in the NL, we obtain $I_{SC}(N) \equiv \sum_{x=0}^{N-1} C(x) = -3/4$, indicating that the SC degree of freedom in the MZM-QD subsystem is fully screened by the NL. It is important to note that both spin and charge clouds contribute to the screening process. Specifically, we can express $I_{SC}(n) = I_{S}(n) + I_{C}(n)$, with $I_{S}(n) \equiv \sum_{x=0}^{n-1} \langle \mathbf{l}_\textrm{QD} \cdot \mathbf{s}_x \rangle$ and $I_{C}(n) = \langle \mathbf{l}_\textrm{QD} \cdot \bm{\eta}_x \rangle$. In Fig.\ref{fig3}(b), we show $I_{C}(N)/I_{S}(N)$ as a function of $T_{A \otimes N}/T_K$. Remarkably, all data collapse onto a universal curve, which increases linearly until it saturates at 0.5. In this sense, $T_{A \otimes N}$ is directly related to the proportion of the charge component in the screening cloud.

As both theoretical and experimental interests, when $U$ changes its sign to a negative value, the effective model remains valid with the model parameters $J^M \to -J^M$ and $J^C \leftrightarrow J^S$, correspondingly. It can be verified that the screening cloud is independent of the sign of $J^M$, while the spin and charge components in the screening cloud exchange roles. In this case, the roles of the $J^C$ and $J^S$ terms are reversed, with the charge cloud dominating the screening process.

\begin{figure}
    \centering
    \includegraphics[width=1.0\linewidth]{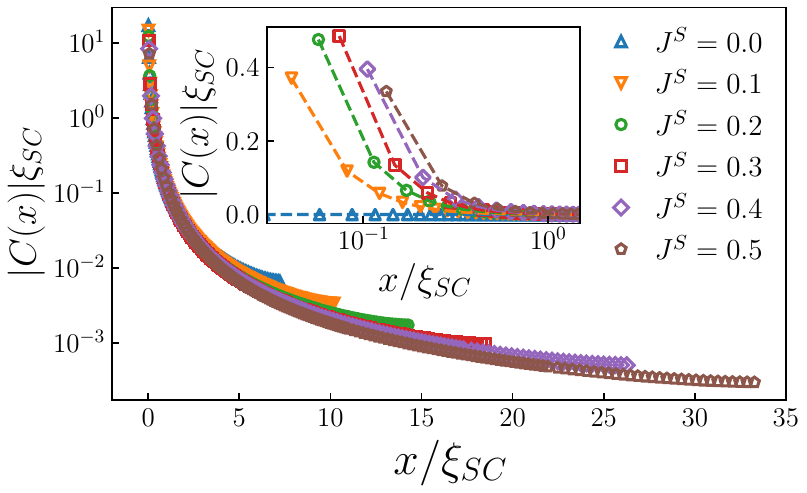}
\caption{Collapse of $C(x)\xi_{SC}$s (for odd $x$s) as functions of $x/\xi$ for various $J^K$ with $J^M=0.3$ and $N = 499$. Inset: Corresponding $C(x)\xi_{SC}$s for even $x$s.}
    \label{fig4}
\end{figure}

\textcolor{blue}{\textit{Scaling of the SC cloud}.--}We now turn to the microscopic structure of the SC screening cloud by investigating $C(x)$ with the NL modeled as a one-dimensional tight-binding chain of length $N$. By exploiting the underlying $\mathbb{Z}_2$ and $SU_L(2)$ symmetries in DMRG, we can accurately obtain $C(x)$ for long chains up to $N = 499$. Defining $I_{SC}(n) \equiv \sum_{x=0}^{n-1} C(x)$, i.e., integrating $C(x)$ up to site $n$, we first determine the characteristic length by defining $\xi_{SC} = \mathrm{min} \{ n : (1 + I_{SC}(n)/0.75) < 0.1 \}$, i.e., the minimal length at which $90\%$ of the SC degree of freedom in the MZM-QD subsystem is screened \cite{Holzner2009}.

In Fig.\ref{fig4}, we show $|C(x)| \xi_{SC}$ as a function of $x/\xi_{SC}$, with $x$ corresponding to odd sites, for various $J^S$, while fixing $J^M = 0.3$. All data collapse onto a universal curve, even for $J^S = 0$ or $J^S \gg J^M$. This suggests that $\xi_{SC}$ is the only characteristic length for the SC cloud. Given that $\xi_{\textrm{sc}}(J^M = 0) \sim 1/T_K$ for the spin Kondo system, we expect $\xi_{SC} \sim 1/T_{SC}$ when $J^M \neq 0$. The sub-leading structure, with $x$ corresponding to even sites, is, however, delicately modified by the $A \otimes N$ boundary condition. In the limit of $J^S = 0$, $|C(x)| \xi_{SC} = 0$ for all even $x$. This unique even-odd behavior is crucial for determining the SC cloud in experiments.

\textcolor{blue}{\textit{Summary}.--}
In summary, by carefully comparing and analyzing the spectral function, thermodynamic properties, and screening cloud using NRG and DMRG, we identify the low-energy physics of the MZM-QD-NL system as a novel SC Kondo effect. The charge degree of freedom is preserved by treating the MZM-QD subsystem as the atomic limit. At the symmetric point, the low-energy levels of the MZM-QD subsystem form two topologically degenerate $SU_L(2)$ singlets, which are further screened by the SC cloud in the NL after hybridization. Notably, the screening process is compatible with the $A \otimes N$ boundary conditions, with the characteristic temperature $T_{A \otimes N}$ of the $A \otimes N$ fixed point closely related to the proportion of the charge component in the screening cloud. At the same time, the leading term in the SC cloud exhibits a similar scaling behavior to the traditional spin Kondo effect, with the characteristic temperature modified from $T_K$ to $T_{SC}$ as $T_{A \otimes N} \gtrsim T_K$. The sub-leading even-odd behavior, however, is delicately modified by the $A \otimes N$ boundary conditions.

Currently, we use the $SU_L(2)$ symmetry to facilitate the underlying screening process. It is expected that the SC cloud will largely persist under $SU_L(2)$-breaking perturbations, provided their strength is much smaller than $T_{SC}$. However, the competition between these perturbations and the SC cloud warrants further investigation. Building on recent advances in anyonic and Kondo systems, such as $\mathbb{Z}_N$ parafermions \cite{Alicea2015,Hutter2015,Alicea2016} and $SU(N)$ Kondo systems \cite{Carmi2011,Moca2012,Lindner2018}, a natural extension of this work is to explore how anyon-Kondo interactions evolve in these new frameworks. This could reveal novel phenomena in strongly correlated systems and topological quantum matter, with potential implications for quantum computation.


\begin{acknowledgments}
We are grateful to Rui Wang, Lihui Pan and Yun Chen for fruitful discussions. This work was supported by the Natural Science Foundation of China (No. 11904245) and the Postdoctoral Science Foundation of China (No. 2021M690330) and was also supported by the Sichuan Normal University High Performance Computing Center, China.

\end{acknowledgments}

\bibliography{references}

\end{document}


\beginsupplement

\title{Supplementary Material for: Spin-charge Kondo effect for a quantum dot with side coupled Majorana zero mode}
\author{Haojie Shen}
\affiliation{School of Physics and Astronomy, Shanghai Jiao Tong University, Shanghai 200240, China}

\author{Wei Su}
\email{suwei@sicnu.edu.cn}
\affiliation{College of Physics and Electronic Engineering, Center for Computational Sciences, Sichuan Normal University, Chengdu 610068, China}
\affiliation{Beijing Computational Science Research Center, Beijing 100084, China}

\author{Mengnan Chen}
\affiliation{School of Science, Hangzhou Dianzi University, Hangzhou 310027, China}

\author{Xiaoqun Wang}
\email{xiaoqunwang@sjtu.edu.cn}
\affiliation{School of Physics, Zhejiang University, Hangzhou 310027, China}

\date{\today}


\maketitle


\appendix
\section{Conservation of Spin-charge rotation symmetry}
\label{sec:appendixA}
At the particle-hole symmetric point, the QD-NL system has both global spin rotation $SU_S(2)$ symmetry and charge rotation $SU_C(2)$ symmetry. The polarized MZM couples exclusively to one component of the spin at the QD, thereby breaking both symmetries. However, we demonstrate that the global spin-charge rotation symmetry $SU_L(2)$ remains preserved. The $SU_L(2)$ symmetry is generated by
\begin{equation}
   \mathbf{L} \equiv \sum_i \mathbf{l}_i \equiv \sum_i \mathbf{s}_i + \bm{\eta}_i,
\end{equation}
where $\mathbf{s}_i$ and $\bm{\eta}_i$ are the generators of spin and charge rotation symmetries at site $i \in \{\textrm{QD}, \textrm{NL}\}$, respectively.

Without loss of generality, we will prove the $SU_L(2)$ symmetry in a one-dimensional chain, where the site(s) at the QD (NL) are labeled with $i = -1$ (for the QD) and $i = 0, 1, 2, \dots, L-1$ (for the NL). To simplify the notation, we will use both the subscript $\text{QD}$ and $-1$, with $c_{-1}$ and $d$ referring to the same annihilation operator at the QD. The ladder operators are defined as:
\begin{equation}
    s^z_i = \frac{n_{i, \uparrow} - n_{i, \downarrow}}{2}, \quad s^+ = c_{i, \uparrow}^{\dagger} c_{i, \downarrow}, \quad s^- = c_{i, \downarrow}^{\dagger} c_{i, \uparrow},
\end{equation}
and
\begin{equation}
    \eta^z_i = \frac{1 - \sum_{\sigma} c^\dagger_{i, \sigma} c_{i, \sigma}}{2}, \quad \eta_i^+ = (-1)^{i} (c_{i \downarrow} c_{i \uparrow}), \quad \eta_i^- = (-1)^{i} (c_{i \uparrow}^\dagger c_{i \downarrow}^\dagger).
\end{equation}

For the Anderson-type Hamiltonian, we require that both $\mathbf{s}_{-1} \equiv (s^x_{-1}, s^y_{-1}, s^z_{-1})$ and $\bm{\eta}_{-1} \equiv (\eta^x_{-1}, \eta^y_{-1}, \eta^z_{-1})$ at the QD commute with the MZM-QD coupling $\gamma (d_{\uparrow} - d_{\uparrow}^{\dagger})$, which can be easily proven through some operator algebra. Therefore, the global $SU_L(2)$ symmetry is conserved. It is important to note that there is a global sign arbitrariness in the definition of $\eta^\pm_i$ in the absence of the MZM, which is later fixed by the MZM coupling.

The conservation of the $SU(2)_L$ symmetry for $H^J_{\textrm{eff}}$ (see main text) can also be demonstrated by showing that
\begin{equation}
   \left[l^\alpha_{\textrm{QD}} + l^\alpha_{0}, \mathbf{l}_\textrm{QD} \cdot \bm{\xi}_0 \right] = 0 \quad (\alpha = x, y, z). 
\end{equation}
Here, we provide an alternative proof by explicitly using the Majorana representation, which is more comprehensive and much easier for generalization. To be consistent with the alternating sign in the definition of $\eta^\pm_i$, we define the Majorana fermion operators as follows:
\begin{equation}
    \bm{\xi}_i = \left[\xi^{x}_{i}, \xi^{y}_{i}, \xi^{z}_{i} \right] \equiv  
    \left\{
    \begin{array}{ll}
        \left[\mathrm{i}( c_{i \downarrow}^{\dagger}-c_{i \downarrow}),  -(c_{i \downarrow}^{\dagger} + c_{i \downarrow}) ,\mathrm{i}( c_{i \uparrow}^{\dagger}- c_{i \uparrow} ) \right], & \text{for even $i$}; \\
        \\[0.1pt] 
        \left[-( c_{i \downarrow}^{\dagger}+c_{i \downarrow}),  \mathrm{i}(-c_{i \downarrow}^{\dagger} + c_{i \downarrow}) ,-( c_{i \uparrow}^{\dagger}+ c_{i \uparrow} ) \right], & \text{for odd $i$}.
    \end{array}
    \right.
\end{equation}
One can show that $\xi^{\alpha}_{i}$($\alpha=x,y,z$) satisfy the Clifford algebra $\left\{\xi_{i}^{\alpha}, \xi_{j}^{\beta}\right\}  = 2 \delta_{\alpha \beta}\delta_{ij}$. The Majorana representation of generator $l_i^\alpha$ is then given by
\begin{equation}
    l^{\alpha}_{i}=\frac{-\mathrm{i}\epsilon_{\alpha\beta\gamma}}{4}\xi_i^{\beta}\xi_i^{\gamma}, 
\end{equation}
where \(\epsilon_{\alpha\beta\gamma}\) is the Levi-Civita symbol. One can prove that
\begin{equation}
    \begin{aligned}
               \left[l_{i}^{\alpha}, \xi_{i}^{\beta}\right]&=\frac{-\mathrm{i} \epsilon_{\alpha \mu \nu}}{4}\left[ \xi_{i}^{\mu} \xi_{i}^{\nu}, \xi_{i}^{\beta}\right]\\
        &=\frac{\mathrm{i} \epsilon_{\alpha \mu \nu}}{4}\left(2\delta_{\mu\beta}\xi^{\nu}_{i}-2\delta_{\nu\beta}\xi^{\mu}_{i}\right)=\mathrm{i} \epsilon_{\alpha \beta \gamma} \xi_{i}^{\gamma}.
    \end{aligned}\label{eq2}
\end{equation}
This shows that $\xi^{\alpha}_{i}$ transformer under the fundamental representation of $SO(3)$. Finally, one has
\begin{equation}
    \begin{aligned} &[l_\textrm{QD}^{\alpha}+l_{0}^{\alpha},l_\textrm{QD}^{\beta}\xi_{0}^{\beta}]\\
    =& l_\textrm{QD}^\beta\left[l_{0}^{\alpha}, \xi_{0}^\beta\right]+\left[l_{\textrm{QD}}^\alpha , l_\textrm{QD}^\beta\right] \xi_{0}^\beta \\
 =&l_\textrm{QD}^\beta \mathrm{i} \epsilon_{\alpha \beta \gamma} \xi_{0}^\gamma+\mathrm{i} \epsilon_{\alpha \beta \gamma} l_\textrm{QD}^\gamma \xi_{0}^\beta=0.
    \end{aligned}
\end{equation}
Q.E.D.

\section{Spectrum for the MZM-QD subsystem}
The full spectrum of the MZM-QD subsystem has already been given in the supplementary material of \cite{Lee2013}. Here, we present the compact form at the particle-hole symmetric point $U = -2\epsilon_d$. The total Hilbert space has dimension 8, with the high- and low-energy states well separated by an energy gap $\Delta_E = \sqrt{\epsilon_d^2 + 4 \lambda^2}$. The low-energy states have energy $\frac{1}{2}(\epsilon_d - \sqrt{\epsilon_d^2 + 4\lambda^2})$ and a degeneracy of 4, which can be grouped by their parity $P = 0, 1$. The two states within each group form an $SU(2)_L$ doublet. Labeling the states as $|P, l_z\rangle$, the wave functions are given by
\begin{align}
    |1,\frac{1}{2}\rangle^l &= \frac{1}{\mathcal{N}} \left[(\sqrt{\epsilon_d^2 + 4\lambda^2}-\epsilon_d) |0\rangle\otimes |\uparrow\rangle - 2\lambda |1\rangle\otimes |0\rangle\right]\notag \\
    |1,-\frac{1}{2}\rangle^l  &= \frac{1}{\mathcal{N}} \left[(\sqrt{\epsilon_d^2 + 4\lambda^2}-\epsilon_d) |0\rangle\otimes |\downarrow\rangle + 2\lambda |1\rangle\otimes |\uparrow\downarrow\rangle\right],
\end{align}
and 
\begin{align}
    |0,\frac{1}{2}\rangle^l &= \frac{1}{\mathcal{N}} \left[(\sqrt{\epsilon_d^2 + 4\lambda^2}-\epsilon_d) |1\rangle\otimes |\uparrow\rangle + 2\lambda |0\rangle\otimes |0\rangle\right] \notag \\
    |0,-\frac{1}{2}\rangle^l &= \frac{1}{\mathcal{N}} \left[(\sqrt{\epsilon_d^2 + 4\lambda^2}-\epsilon_d) |1\rangle\otimes |\downarrow\rangle - 2\lambda |0\rangle\otimes |\uparrow\downarrow\rangle\right].
\end{align}
The high-energy states have energy $\frac{1}{2}(\epsilon_d +\sqrt{\epsilon^2+4\lambda^2})$. The wave functions are
\begin{align}
    |1,\frac{1}{2}\rangle^h &= \frac{1}{\mathcal{N}} \left[2\lambda |0\rangle\otimes |\uparrow\rangle +(\sqrt{\epsilon_d^2 + 4\lambda^2}-\epsilon_d) |1\rangle\otimes |0\rangle\right] \notag\\
    |1,-\frac{1}{2}\rangle^h &= \frac{1}{\mathcal{N}} \left[2\lambda |0\rangle\otimes |\downarrow\rangle - (\sqrt{\epsilon_d^2 + 4\lambda^2}-\epsilon_d) |1\rangle\otimes |\uparrow\downarrow\rangle\right],
\end{align}
and
\begin{align}
    |0,\frac{1}{2}\rangle^h &= \frac{1}{\mathcal{N}} \left[2\lambda |1\rangle\otimes |\uparrow\rangle - (\sqrt{\epsilon_d^2 + 4\lambda^2}-\epsilon_d) |0\rangle\otimes |0\rangle\right] \notag\\
    |0,-\frac{1}{2}\rangle^h &= \frac{1}{\mathcal{N}} \left[2\lambda |1\rangle\otimes |\downarrow\rangle + (\sqrt{\epsilon_d^2 + 4\lambda^2}-\epsilon_d) |0\rangle\otimes |\uparrow\downarrow\rangle\right].  
\end{align}
With $\mathcal{N}$ the corresponding normalization constant.

\section{Green's functions from the equation of motion}
Since the Hubbard $U$ operates only on the QD, Green's functions at the MZM or in the NL can be deduced from those at the QD. Here, we present the deduction using the standard equation of motion technique, within which we have
\begin{equation}
\begin{aligned}
    &\omega\langle\langle A | B \rangle\rangle_\omega = \langle [A, B]_+\rangle + \langle\langle [A,H] | B \rangle\rangle_\omega\\
    &\omega\langle\langle A | B \rangle\rangle_\omega = \langle [A, B]_+\rangle - \langle\langle A | [B,H] \rangle\rangle_\omega,
\end{aligned}
\end{equation}
Here, $\langle\langle A | B \rangle\rangle_\omega$ denotes the Green's function of two operators $A$ and $B$ in the frequency domain. Using the Hamiltonian (1) defined in the main text, we obtain
\begin{equation}
\begin{aligned}
    \omega \langle\langle c_\mathbf{k}|c_{\mathbf{k}^\prime}^\dagger \rangle\rangle &= \delta_{\mathbf{k}, \mathbf{k}^\prime} + \epsilon_{\mathbf{k}} \langle\langle c_\mathbf{k}|c^\dagger_{\mathbf{k}^\prime} \rangle\rangle +V_\mathbf{k}\langle\langle d| c_{\mathbf{k}^\prime}^\dagger \rangle\rangle \\
    \omega \langle\langle d|c_{\mathbf{k}^\prime }^\dagger\rangle\rangle &= \epsilon_{\mathbf{k}^\prime} \langle\langle d|c_{\mathbf{k}^\prime }^\dagger \rangle\rangle +V_{\mathbf{k}^\prime}\langle\langle d|d^\dagger\rangle\rangle \\
    \omega \langle\langle c^\dagger_\mathbf{k}|c_{\mathbf{k}^\prime}^\dagger \rangle\rangle &= -\epsilon_{\mathbf{k}} \langle\langle c^\dagger_\mathbf{k}|c^\dagger_{\mathbf{k}^\prime} \rangle\rangle -V_\mathbf{k}\langle\langle d^\dagger| c_{\mathbf{k}^\prime}^\dagger \rangle\rangle \\
    \omega \langle\langle d^\dagger|c_{\mathbf{k}^\prime }^\dagger\rangle\rangle &= \epsilon_{\mathbf{k}^\prime} \langle\langle d^\dagger|c_{\mathbf{k}^\prime }^\dagger \rangle\rangle +V_{\mathbf{k}^\prime}\langle\langle d^\dagger|d^\dagger\rangle\rangle,
\end{aligned}
\end{equation}
Here, we consider only the spin-$\uparrow$ component, and the subscript $\uparrow$ for the spin is omitted. By eliminating $\langle\langle d | c_{\mathbf{k}^\prime}^\dagger \rangle\rangle$ and $\langle\langle d^\dagger | c_{\mathbf{k}^\prime}^\dagger \rangle\rangle$ from these equations, the Green's function for the conduction electrons takes the compact form
\begin{equation}
    G^{c}=\begin{bmatrix}
        \langle\langle c_{\mathbf{k}}| c_{\mathbf{k}^\prime}^\dagger\rangle\rangle &  \langle\langle c_{\mathbf{k}}| c_{\mathbf{k}^\prime}\rangle\rangle\\
        \langle\langle c_{\mathbf{k}}^\dagger| c_{\mathbf{k}^\prime}^\dagger\rangle\rangle & \langle\langle c_{\mathbf{k}}^\dagger| c_{\mathbf{k}^\prime}\rangle\rangle
    \end{bmatrix}  = 
    G_0^{c}+ G_0^{c} T G_0^{c}, \label{Gc}
\end{equation}
Here
\begin{equation}
    G_0^{c} = \begin{bmatrix}
        \langle\langle c_{\mathbf{k}}| c_{\mathbf{k}^\prime}^\dagger\rangle\rangle_0 &         \\
        &         \langle\langle c_{\mathbf{k}}^\dagger| c_{\mathbf{k}^\prime}\rangle\rangle_0
    \end{bmatrix} =
    \begin{bmatrix}
        \frac{\delta_{\mathbf{k}\mathbf{k}^\prime}}{\omega - \epsilon_{\mathbf{k}}} &         \\
        &         \frac{\delta_{\mathbf{k}\mathbf{k}^\prime}}{\omega + \epsilon_{\mathbf{k}}}
    \end{bmatrix}, 
\end{equation}
is the free Green's function for electrons in the NL, and the T-matrix is given by
\begin{equation}
    T = V_{\mathbf{k}}^\ast
    \begin{bmatrix}
        \langle\langle d| d^\dagger\rangle\rangle& -\langle\langle d| d\rangle\rangle\\
        -\langle\langle d^\dagger| d^\dagger\rangle\rangle&\langle\langle d^\dagger| d\rangle\rangle
    \end{bmatrix}
    V_{\mathbf{k}^\prime}.
\end{equation}

Similarly, the Green's function at the MZM in terms of $f$ is given by
\begin{equation}
    G^{f} = \begin{bmatrix}
        \langle\langle f| f^\dagger\rangle\rangle &  \langle\langle f| f\rangle\rangle\\
        \langle\langle f^\dagger| f^\dagger\rangle\rangle & \langle\langle f^\dagger| f\rangle\rangle
    \end{bmatrix}  = 
    G_0^f+ G_0^f T^f G_0^f, \label{GM}
\end{equation}
where
\begin{equation}
    G_0^f = \begin{bmatrix}
        \frac{1}{\omega+i\eta} &         \\
        &         \frac{1}{\omega+i\eta }
    \end{bmatrix}, 
\end{equation}
and
\begin{equation}
    T^f = \begin{bmatrix}
        \lambda & \lambda         \\
        \lambda &  \lambda
    \end{bmatrix} 
    \begin{bmatrix}
        \langle\langle d| d^\dagger\rangle\rangle& -\langle\langle d| d\rangle\rangle\\
        -\langle\langle d^\dagger| d^\dagger\rangle\rangle&\langle\langle d^\dagger| d\rangle\rangle
    \end{bmatrix} 
    \begin{bmatrix}
        \lambda & \lambda         \\
        \lambda & \lambda
    \end{bmatrix}. 
\end{equation}
Apparently, the anomalous component $\langle\langle d_\uparrow | d_\uparrow \rangle\rangle$ cannot be neglected when extracting spectral information from the MZM, QD, or NL.

In the absence of Hubbard $U$, the Green's functions in the equation of motion are closed, i.e.,
\begin{equation}
\begin{aligned}
    &\left[\omega-\epsilon_d - \sum_\mathbf{k}\frac{V_\mathbf{k}^2}{\omega-\epsilon_\mathbf{k}}\right]\langle\langle d | d^\dagger\rangle\rangle = 1
 - \lambda \left( \langle\langle f| d^\dagger\rangle\rangle +\langle\langle f^\dagger| d^\dagger\rangle\rangle \right) \\
    &\left[\omega+\epsilon_d - \sum_\mathbf{k}\frac{V_\mathbf{k}^2}{\omega+\epsilon_\mathbf{k}}\right]\langle\langle d^\dagger | d^\dagger\rangle\rangle =  \lambda \left(\langle\langle f^\dagger | d^\dagger\rangle\rangle +  \langle\langle f | d^\dagger\rangle\rangle \right)\\
    &\omega\langle\langle f|d^\dagger \rangle\rangle = - \lambda\langle\langle d|d^\dagger\rangle\rangle + \lambda \langle\langle d^\dagger |d^\dagger\rangle\rangle \\
    &\omega \langle\langle f^\dagger|d^\dagger \rangle\rangle = \lambda\langle\langle d^\dagger|d^\dagger\rangle\rangle - \lambda \langle\langle d |d^\dagger\rangle\rangle
\end{aligned}.\label{EOM}
\end{equation}
Green's functions at the QD can be solved as
\begin{equation}
    \langle\langle d|d^\dagger \rangle\rangle = \frac{1}{\omega-\epsilon_d+ i \Gamma-\frac{2 \lambda^{2}}{\omega-\frac{2 \lambda^{2}}{\omega+\epsilon_d-i \Gamma}}}, 
\end{equation}
and 
\begin{equation}
    \langle \langle d^\dagger|d^\dagger\rangle\rangle=\frac{2 \lambda}{\omega-\frac{2 \lambda^{2}}{\omega+\epsilon_d+i \Gamma}}\frac{-\lambda}{\omega+\epsilon_d+i \Gamma}\langle\langle d|d^\dagger \rangle\rangle. 
\end{equation}

\section{Proof of $\rho_\downarrow(\omega) = \rho_\uparrow(\omega) + A_\uparrow(\omega) $}
Based on the definition of $\bm{\xi}_i$ in Eq. (S5), we obtain $l_{\textrm{QD}}^{\alpha}=\frac{-\mathrm{i}\epsilon_{\alpha\beta\gamma}}{4}\xi_{\textrm{QD}}^{\beta}\xi_{\textrm{QD}}^{\gamma}$, $d_{\downarrow}=\frac{-\mathrm{i}\xi^{y}_{\textrm{QD}}-\xi^{x}_{\textrm{QD}}}{2}$ and
\begin{equation}
    \begin{aligned}
        \left[l_{\textrm{QD}}^{\alpha} \xi_{\textrm{QD}}^{\beta}\right]&=\mathrm{i} \epsilon_{\alpha \beta \gamma} \xi_{\textrm{QD}}^{\gamma}=\xi_{\textrm{QD}}^{\gamma}\left(l_{\textrm{SO(3)}}\right)_{\gamma \beta}^{\alpha},
    \end{aligned}
\end{equation}
where in the last equation, we rewrite $\textrm{i}\epsilon_{\alpha\beta\gamma}$ as the three fundamental representation matrices $\left(l_{\textrm{SO(3)}}\right)_{\gamma \beta}^{\alpha}$ for the $SO(3)$ group. Using the Baker–Campbell–Hausdorff formula, the $SU_L(2)$ rotation of $\bm{\xi}_{\textrm{QD}}$ is then given by

\begin{equation}
e^{\mathrm{i}  \theta \bm{n} \cdot \bm{l}} \xi_{\textrm{QD}}^{\gamma} e^{-\mathrm{i} \theta \bm{n} \cdot \bm{l}} = \left(e^{\mathrm{i}  \theta n_{\alpha} l_{\textrm{SO(3)}}^{\alpha}}\right)_{\beta \gamma} \xi_{\textrm{QD}}^{\beta},
\end{equation}
where $\theta$ is the rotation angle and $\bm{n} = (n_x, n_y, n_z)$ is the unit vector along the rotation axis. The rotation matrix is

\begin{equation}
e^{\mathrm{i}  \theta n^{\alpha} l_{\textrm{SO(3)}}^{\alpha}}
=\left(\begin{array}{lll}
2\left(n_{x}^{2}-1\right) s^{2}+1 & 2 n_{x}n_{y} s^{2}+2 n_{z}cs & 2 n_{x} n_{z} s^{2}-2 n_{y} c s \\
2 n_{x} n_{y} s^{2}-2 n_{z} c s & 2\left(n_{y}^{2}-1\right) s^{2}+1 & 2 n_{y} n_{z} s^{2}+2 n_{x} c s \\
2 n_{x} n_{z} s^{2}+2 n_{y} c s & 2 n_{y} n_{z} s^{2}-2 n_{x} c s & 2\left(n_{z}^{2}-1\right) s^{2}+1
\end{array}\right),
\end{equation}
where $c\equiv \cos \frac{\theta}{2}$ and $s\equiv \sin \frac{\theta}{2}$. The $SU_L(2)$ rotation for $d_{\downarrow}$ is thus
\begin{eqnarray}
e^{\mathrm{i}  \theta \bm{n} \cdot \bm{l}} d_{ \downarrow} e^{-\mathrm{i} \theta \bm{n} \cdot \bm{l}}
= && -\frac{1}{2}\left[\left(2\left(n_{x}^{2}-1\right) s^{2}+1\right) \xi^{x}_{\textrm{QD}} +\left(2 n_{x}  n_{y} s^{2}-2 n_{z} c s\right) \xi^{y}_{\textrm{QD}}+\left(2 n_{x} n_{z} s^{2}+2 n_{y} c s\right) \xi^{z}_{\textrm{QD}}\right] \nonumber\\
&& - \frac{\mathrm{i}}{2} \left[\left(2 n_{x} n_{y} s^{2}+2 n_{z}cs \right)\xi^{x}_{\textrm{QD}}+\left(2\left(n_y^{2}-1\right) s^{2}+1\right) \xi^{y}_{\textrm{QD}} +\left(2 n_{y} n_{z}s^{2}-2 n_{x} c s\right) \xi^{z}_{\textrm{QD}}\right].
\end{eqnarray}
By setting $n_z = n_x = 0$, $n_y = 1$, and $\theta = \frac{\pi}{2}$, $d_{\downarrow}$ is rotated to 
\begin{equation}
 \frac{-\mathrm{i}\xi^{y}_{\textrm{QD}}-\xi^{z}_{\textrm{QD}}}{2} =\frac{1}{2}\left(-d_{\downarrow}^{\dagger} + d_{\downarrow}+ d_{\uparrow}^{\dagger} + d_{\uparrow} \right).
\end{equation}
Owing to the global $SU_2(2)$ symmetry of the MZM-QD-NL system, it follows that
\begin{eqnarray}
\left\langle \left\langle d_{\downarrow} \left| d_{\downarrow}^{\dagger} \right\rangle \right\rangle_{\omega}\right. &=&
\frac{1}{4}\left\langle \left\langle \left(-d_{\downarrow}^{\dagger} + d_{\downarrow}+ d_{\uparrow}^{\dagger} + d_{\uparrow} \right) \left| \left(-d_{\downarrow} + d_{\downarrow}^{\dagger} + d_{\uparrow} + d_{\uparrow}^{\dagger} \right) \right\rangle \right\rangle_{\omega}\right. \nonumber\\
&=&\frac{1}{2}\left[ \left\langle \left\langle d_{\downarrow} \left| d_{\downarrow}^{\dagger} \right\rangle \right\rangle_{\omega}\right. 
+ \left\langle \left\langle d_{\uparrow} \left| d_{\uparrow}^{\dagger} \right\rangle \right\rangle_{\omega}\right. 
+ \left\langle \left\langle d^\dagger_{\uparrow} \left| d_{\uparrow}^{\dagger} \right\rangle \right\rangle_{\omega}\right. 
\right],
\end{eqnarray}
where we have used
\begin{eqnarray}
\left\langle \left\langle d_{\sigma} \mid d_{\sigma}^{\dagger}\right\rangle\right\rangle&=&\left\langle\left\langle d_{\sigma}^{\dagger} \mid d_{\sigma}\right\rangle\right\rangle \nonumber \\
\left\langle \left\langle d_{\sigma} \mid d_{\sigma}\right\rangle\right\rangle&=&\left\langle\left\langle d_{\sigma}^{\dagger} \mid d_{\sigma}^{\dagger}\right\rangle\right\rangle,
\end{eqnarray}
due to the particle-hole symmetry.

The normal and anomalous LDOS are defined by
\begin{equation}
\rho_\sigma(\omega)\equiv -\frac{1}{\pi} \textrm{Im} \left\langle \left\langle d_{\sigma} \left| d_{\sigma}^{\dagger} \right\rangle \right\rangle_{\omega}\right.,
\end{equation}
\begin{equation}
A_\uparrow(\omega)\equiv -\frac{1}{\pi} \textrm{Im} \left\langle \left\langle d^\dagger_{\uparrow} \left| d_{\uparrow}^{\dagger} \right\rangle \right\rangle_{\omega}\right.,
\end{equation}
respectively. We finally derive the relation of $\rho_{\downarrow}(\omega)=\rho_{\uparrow}(\omega)+A_{\uparrow}(\omega)$.

\section{Effective low energy Hamiltonian}
When $V_\mathbf{k} \ll \Delta_E = \sqrt{\epsilon_d^2 + 4 \lambda^2}$, i.e., when the hybridization is much smaller than the energy gap, we can apply the Schrieffer-Wolff transformation to project out the high-energy states. The effective low-energy Hamiltonian can then be formally expressed as

\begin{equation}
    H_{\mathrm{eff}}=H_{ll}+\sum_{h} H_{l h} \frac{1}{E-H_{h h}} H_{h l},
\end{equation}
Where $H_{mn} = P_{m} H P_{n}$, and $P_{n} = |n\rangle \langle n|$ is the projection operator, with $n = h$ ($l$) denoting high (low) eigenstates of the MZM-QD subsystem (see Eq. (S9)–(S12)). Since both the low and high-energy states are superpositions of states with different occupation numbers, it is difficult to explicitly write down the projection operator $P$ using local operators. Alternatively, we perform the Schrieffer-Wolff transformation by directly evaluating the inner products in $H_{mn} = P_{m} H P_{n} = |n\rangle \langle n| H |m\rangle \langle m|$.

$H_{ll}$ results from the “real processes,” i.e., transitions within the low-energy states. Explicitly writing this out,
\begin{widetext}
   \begin{equation}
H_{ll}=\left(|1,\frac{1}{2}\rangle^{l},|0,\frac{1}{2}\rangle^{l},|1,-\frac{1}{2}\rangle^{l},|0,-\frac{1}{2}\rangle^{l}\right)
\begin{pmatrix} 
0 & \alpha\left(c_{\mathbf{k\uparrow}}-c_{\mathbf{k\uparrow}}^{\dagger}\right) & 0 & -2\alpha c_{\mathbf{k\downarrow}}^{\dagger}\\
\alpha\left(c_{\mathbf{k\uparrow}}^{\dagger}-c_{\mathbf{k\uparrow}}\right) & 0 & 2\alpha c_{\mathbf{k\downarrow}}^{\dagger} & 0\\
0 & 2\alpha c_{\mathbf{k\downarrow}} & 0 & \alpha\left(c_{\mathbf{k\uparrow}}^{\dagger}-c_{\mathbf{k\uparrow}}\right)\\
-2\alpha c_{\mathbf{k\downarrow}} & 0 & \alpha\left(c_{\mathbf{k\uparrow}}-c_{\mathbf{k\uparrow}}^{\dagger}\right) & 0
\end{pmatrix}
\left(\begin{array}{c}
{}^l\langle1,\frac{1}{2}|\\
{}^l\langle0,\frac{1}{2}|\\
{}^l\langle1,-\frac{1}{2}|\\
{}^l\langle0,-\frac{1}{2}|
\end{array}\right).
\end{equation} 
\end{widetext}
Here, the summation over $\mathbf{k}$ is implicitly understood. Since the low-energy states form two groups of $SU_L(2)$ doublets, one has
\begin{align}
    l_{\textrm{QD}}^{z}|1(0),\pm\frac{1}{2}\rangle^{l} &= \pm\frac{1}{2}|1(0),\pm\frac{1}{2}\rangle^{l},
\end{align}
and
\begin{align}
    &l_{\textrm{QD}}^-|1(0),\frac{1}{2}\rangle^{l} = |1(0),-\frac{1}{2}\rangle^{l}\notag\\
    &l_{\textrm{QD}}^+|1(0),-\frac{1}{2}\rangle^{l} = |1(0),\frac{1}{2}\rangle^{l}.
\end{align}
Using these operators, $H_{ll}$ can be compactly written as
\begin{eqnarray}
H_{ll} &=& J^M \left[(\gamma l_{\textrm{QD}}^{z}/\theta)(c_{\mathbf{k}\uparrow}-c_{\mathbf{k}\mathbf{\uparrow}}^{\dagger})+(\gamma l_{\textrm{QD}}^{+}/\theta)(-c_{\mathbf{k}\downarrow}^{\dagger})+(\gamma l_{\textrm{QD}}^{-}/\theta)c_{\mathbf{k}\downarrow} \right]\nonumber\\
    &=&J^M \left[(\gamma l_{\textrm{QD}}^{z}/\theta)(c_{0\uparrow}-c_{0\mathbf{\uparrow}}^{\dagger})+(\gamma l_{\textrm{QD}}^{+}/\theta)(-c_{0\downarrow}^{\dagger})+(\gamma l_{\textrm{QD}}^{-}/\theta)c_{0\downarrow} \right]\nonumber\\
    &=&J^M (\gamma\mathbf{l_{\textrm{QD}}}/\theta)\cdot\bm{\xi}_{0},
\end{eqnarray}
where 
\begin{equation}
\theta=-\frac{\epsilon_{d}}{\sqrt{\epsilon_{d}^{2}+4\lambda^{2}}}.
\end{equation}
In the last equality of Eq. (S37), we have assumed a constant hybridization $V_\mathbf{K} = V$, and the subscript $0$ denotes the origin in the NL that couples to the QD. In this limit
\begin{equation}
    J^{M}=2\alpha=\frac{2\lambda V}{\sqrt{\epsilon_{d}^{2}+4\lambda^{2}}}\xrightarrow{\lambda\to0} \frac{2\lambda V}{|\epsilon_d|},
\end{equation}
which is of first order in $V$.

The second term in Eq. (S43) arises from ``virtual processes,'' i.e., processes mediated by instantaneous transitions to higher energy states. Retaining only contributions up to order $O(V^2)$, we have

\begin{equation}
    \sum_{h} H_{l h} \frac{1}{E-H_{h h}} H_{h l} = J^S\mathbf{l}_{\textrm{QD}}\cdot\mathbf{s}_{0}+J^{C}\mathbf{l}_{\textrm{QD}}\cdot\bm{\eta}_{0} + H_{\textrm{NL}},
\end{equation}
where
\begin{equation}
\begin{aligned}
    J^{S} &= 2V^{2}\frac{\epsilon_{d}(\epsilon_{d}-\sqrt{\epsilon_{d}^{2}+4\lambda^{2}})}{\left(\epsilon_{d}^{2}+4\lambda^{2}\right)^{3/2}}  \xrightarrow{\lambda\to0} 2V^2 \left(-\frac{1}{\epsilon_{d}}+\frac{1}{|\epsilon_{d}|}\right),
\end{aligned}
\end{equation}

\begin{equation}
\begin{aligned}
J^{C} &= 2V^{2}\frac{\epsilon_{d}(\epsilon_{d}+\sqrt{\epsilon_{d}^{2}+4\lambda^{2}})}{\left(\epsilon_{d}^{2}+4\lambda^{2}\right)^{3/2}}\xrightarrow{\lambda\to0} 2V^2 \left(\frac{1}{\epsilon_{d}}+\frac{1}{|\epsilon_{d}|}\right).
\end{aligned}
\end{equation}

In summary, by projecting out the high-energy states, we obtain the effective low-energy coupling $H^J_{\textrm{eff}}$ between the MZM-QD subsystem and the NL. In the constant hybridization limit,
\begin{equation}
H^J_{\textrm{eff}}=J^M(\mathrm{i}\gamma\mathbf{l}_{\textrm{QD}}/\theta)\cdot\bm{\xi}_{0}+J^S\mathbf{l}_{\textrm{QD}}\cdot\mathbf{s}_{0}+J^{C}\mathbf{l}_{\textrm{QD}}\cdot\bm{\eta}_{0}.
    \label{eq4}
\end{equation}
In the large positive $U$ limit, the charge freedom is frozen on the QD, $\gamma \mathbf{l_{\textrm{QD}}}/\theta \to \gamma \mathbf{s}_{\textrm{QD}}$, $\mathbf{l_{\textrm{QD}}}\to \mathbf{s}_{\textrm{QD}}$, and the $J^Q$ term vanished.

\section{Numerical Details}
We use a discretization parameter $\Lambda = 2.5$ and retain $D \sim 2048$ states during the NRG iteration. The $Z_2 \times U_{L}(1)$ symmetry is employed for acceleration, and convergence with respect to $D$ is carefully assumed. The spectral functions are computed within the FDM framework, where a Log-Gaussian function with parameter $b = 0.9$ is used for broadening. In calculating $\chi_{C}$ and $\chi_{S}$, we introduce a small $h l_z$ term to the QD with $h = 2 \times 10^{-6}$, and the derivative is calculated discretely.

We employed the $Z_2 \times SU_L(2)$ symmetry in DMRG calculations, with a maximum bond dimension $D \sim 1000$ (effectively $D \sim 3400$ for $U_{L}(1)$ symmetry). The truncation error is $\sim 2 \times 10^{-8}$ for the largest system with $N = 499$. The use of $SU_L(2)$ symmetry is crucial for accurately determining the correlation function $\langle \mathbf{l}_{\textrm{QD}} \cdot \mathbf{l}_x \rangle$. The character length $\xi_{SC}$ of the SC cloud is determined in the same manner as the Kondo cloud in Ref. \cite{Holzner2009}. Here, we generalize the definition of the integrated correlation function to

 \begin{equation}
\Sigma(n)=1+\sum_{x=0}^{n-1} \frac{\left\langle\mathbf{l}_\textrm{QD} \cdot\bm{\mathbf{l}}_x\right\rangle}{\left\langle\mathbf{l}_\textrm{QD} \cdot \mathbf{l}_\textrm{QD}\right\rangle}=1+I_{SC}(n)/0.75
\end{equation}
For sufficiently large $n$, one has $I_{SC}(n) \equiv \sum_{x=0}^{n-1} \langle \mathbf{l}_{\textrm{QD}} \cdot \mathbf{l}_x \rangle = -0.75$ (or $\Sigma(n) = 0$), indicating that the local SC degree of freedom at the MZM-QD subsystem is fully screened by those in the NL. The screening length $\xi_{SC}$ is extracted from $\Sigma(n)$, defined as the length scale at which a certain fraction $a$ of the SC degree of freedom is screened, i.e.

\begin{equation}
\xi_a(N)=\min \{n ; \Sigma(n) \leq 1-a\}.
\end{equation}
In Fig. (S1)(a), we present \( \Sigma(n) \) for various system sizes \( N \). The calculation uses \( a = 90\% \), as indicated by the dashed horizontal line. It is evident that the calculated \( \xi_{SC} \) depends on \( N \) and only becomes convergent when \( N \gg \xi_{SC} \), as shown in Fig. (S1)(b). In the main text, we include only the convergent data for \( N = 499 \).

\begin{figure}[h]
    \centering
    \includegraphics[width=1.0\linewidth]{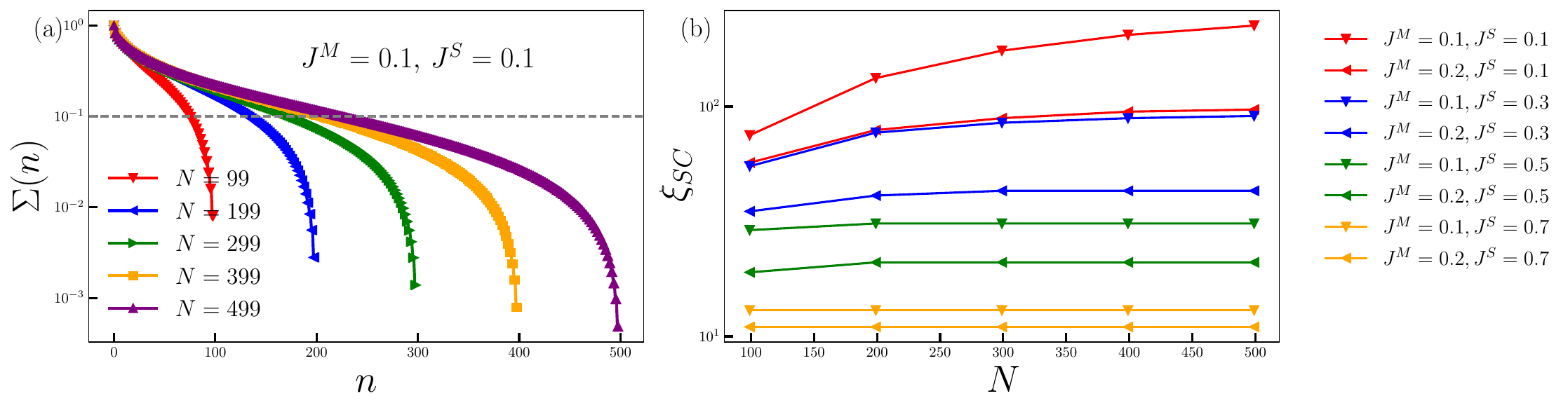}
    \caption{(a) Integrated spin-spin correlations \( \Sigma(n) \) [see Eq. (S44)] for systems of different lengths \( N \). The dashed horizontal line indicates the threshold of 0.1 (or \( a = 90\% \) in Eq. (S45)) used to extract \( \xi_{SC} \). (b) \( \xi_{SC} \) calculated for various values of \( N \) and parameters. It can be seen that \( N = 499 \) is long enough to ensure the convergence of \( \xi_{\textrm{SC}} \), except when both \( J^M \) and \( J^S \) are small (e.g., \( J^M = J^S = 0.1 \)).
}
    \label{smfig1}
\end{figure}

\bibliography{references}